Original Paper

Taichi Murayama[1], Nobuyuki Shimizu[2], Sumio Fujita[2], Shoko Wakamiya[1], Eiji Aramaki[1]

[1] Social Computing Lab, Graduate School of Information Science, Nara Institute of Science and Technology, Ikoma, Japan
[2] Yahoo Japan Corporation, Tokyo, Japan

# Influenza Surveillance using Search Engine, SNS, On-line Shopping, Q&A Service and Past Flu Patients

## Abstract

**Background:** Influenza, an infectious disease, causes many deaths worldwide. Predicting influenza victims during epidemics is an important task for clinical, hospital, and community outbreak preparation.

**Objective:** On-line user-generated contents (UGC), primarily in the form of social media posts or search query logs, are generally used for prediction for reaction to sudden and unusual outbreaks. However, most studies rely only on the UGC as their resource and do not use various UGCs. Our study aims to solve these questions about Influenza prediction: Which model is the best? What combination of multiple UGCs works well? What is the nature of each UGC?

**Methods:** We adapt some models, LASSO Regression, Huber Regression, Support Vector Machine regression with Linear kernel (SVR) and Random Forest, to test the influenza volume prediction in Japan during 2015 – 2018. For that, we use on-line five data resources: (1) past flu patients, (2) SNS (Twitter), (3) search engines (Yahoo! Japan), (4) shopping services (Yahoo! Shopping), and (5) Q&A services (Yahoo! Chiebukuro) as resources of each model. We then validate respective resources contributions using the best model, Huber Regression, with all resources except one resource. Finally, we use Bayesian change point method for ascertaining whether the trend of time series on any resources is reflected in the trend of flu patient count or not.

**Results:** Our experiments show Huber Regression model based on various data resources produces the most accurate results. The coefficient of determination $R^2$ are 0.907 from 2015 – 2016, 0.889 from 2016 – 2017 and 0.917 from 2017 – 2018 in predicting influenza 2 week ahead of the current date. Then, from the change point analysis, we get the result that search query logs and social media posts for three years represents these resources as a good predictor. The results of *Sensitivity*, is one metric, are 80% in search query and 75% in social media posts from 2017–2018.

**Conclusions:** We show that Huber Regression based on various data resources is strong for outliers and is suitable for the flu prediction. Additionally, we indicate the characteristics of each resource for the flu prediction.

## Introduction

### Background

Predicting epidemics of infections can reduce time and effort exerted by public health professionals and medical institutions, resulting in more efficient service. Influenza, especially infections commonly known as the flu, can infect people of any age: even today, it causes 290,000–650,000 deaths annually [1]. In Japan, predicting the dynamics of seasonal flu outbreaks is important. Two well-known aspects of flu outbreaks in temperate areas such as Japan are that (i) flu outbreaks happen every year during December–March and (ii) the flu has a seasonal cycle. Considering these features, one study has predicted outbreaks using only the number of flu patients in the past [2]. However, this research exhibits some limitation especially in term of the method that is used, it can not handle unexpected epidemics that differ from those of a usual year.

To address that shortcoming, on-line user-generated contents (UGC) have become widely used. The meanings of UGC vary among fields. Herein, UGC primarily refers to various on-line services such as social media posts (e.g., Twitter), search query logs (e.g., Google) and on-line shopping records. As new resources to catch and predict flu outbreaks, UGCs are attracting attention. Especially, several UGCs such as Google [3], Twitter [4], and Wikipedia [5] are used. More recently, the manner of UGC use has been refined. One study proposed a mode of selecting queries [6] and a mode of using defective UGC data for economically developing countries [7]. Although various methods exist based on UGC, most of them share one shortcoming: reliance on a single resource. As examples of using various UGC, Santillana et al. [8] used Twitter, Google Trends and flu report. Nevertheless, little is known about comparisons among resources, such as which UGC is the best, which combination of multiple UGCs works well, and the nature of each UGC. Much room exists for the study of multiple UGC handling.

This study uses resources of five kinds; (1) past flu patients, (2) social media posts, (3) search query logs, (4) shopping service query logs, and (5) Q&A service query logs from 2013–2018 (Figure 1). Furthermore, by application of linear and non-linear regression, we predict the number of flu patients in the divided term into thirds between 2015-05-25 and 2018-05-21 and discuss the properties of the respective resources.

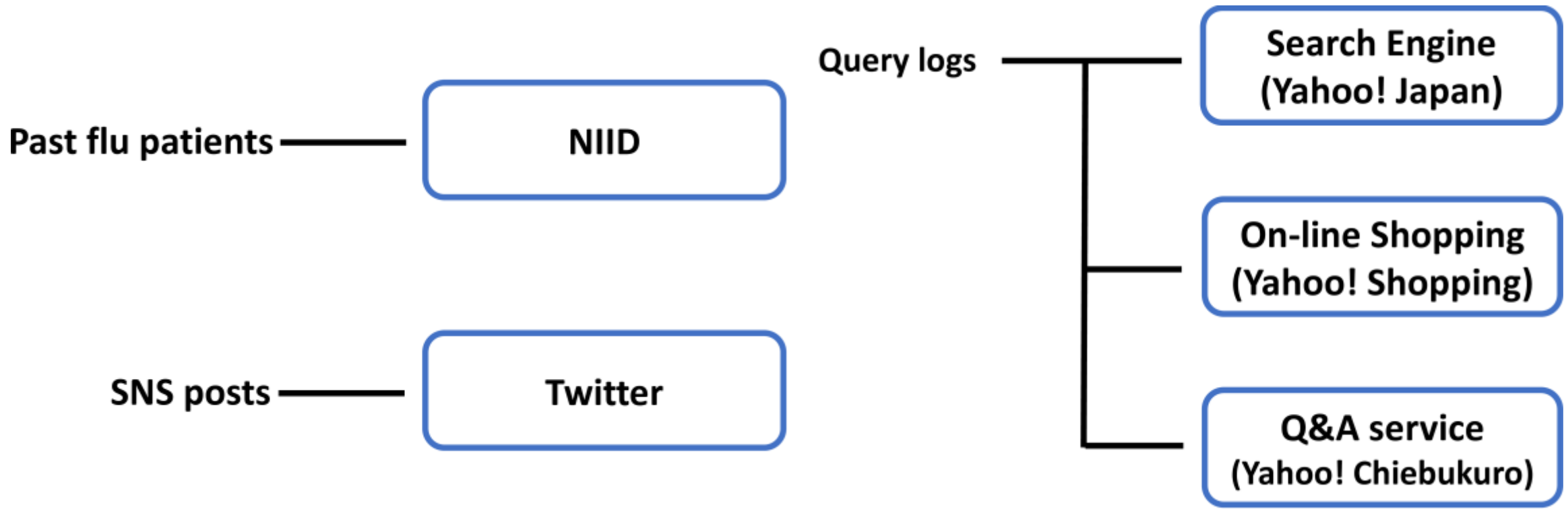

Figure 1. System diagram of using five data resources

This paper's main contributions are the following:

- This flu prediction model incorporates the greatest number of UGCs as predictors.
- Results show that Huber Regression, which is robust against outliers, is suitable for flu prediction using UGC contents.
- Details of the nature of UGCs are discussed.

## Related Work

Studies predicting ILI patient rates/numbers with the flu can be classified into the following three themes: What resources have been used?, How have queries been selected? and What models have been applied?

- What resources have been used?

This question represents one of the important topics in the fields of infodemiology and infoveillance [9]. For the prediction of ILI rates, UGC data over search engines [3, 8, 10, 11, 12] and social networking services [4, 8, 13, 14, 15, 28] have been generally used. Discussions about which service brings more useful resources, search engines or Twitter, have occurred in earlier reports [5, 16, 17]. Aside from UGC data, many studies use various resources for improving models: Wikipedia [5], historical data [2, 10], weather data [18, 19], and self-reporting mobile app [20]. Some studies use various resources together [8, 10, 21, 22]. We make the ensemble model combining five more independent resources than previous studies including query logs about on-line shopping and Q & A service, never been used. Our research then aims to find the characteristic of each resource, which we use when making the influenza prediction model.

- How have queries been selected?

As utilizing UGC data concerning a specific topic over search engines and social networking services, it is an important task to select appropriate queries because the performance would be highly dependent on the task. For selecting suitable queries for infectious diseases, it is general to use a method based on correlation between the ILI rate and the volumes of queries. For example, the method using Google Correlate [23] and the one using word embedding [6] have been developed. Our research uses the commonly used method, correlation between the ILI rate and the volumes of queries.

- What models have been applied?

Prediction of the ILI rate frequently uses linear regression because it is simple model and achieves measurable prediction. Additionally, various models have been proposed to apply flu prediction: the ARIMA model [24], Random Forest [19, 24], disease models such as IDEA model [25], LSTM [22] and Bayesian model [11].

## Methods

### Datasets

For predicting the number of flu patients in a week, we use five resources as features for 2013-10-12 through 2018-05-21. Five resources are past flu patients (National Institute of Infectious Diseases (NIID)), Search Engine (Yahoo! Japan), Social Networking Service (Twitter), On-line shopping service (Yahoo! Shopping (Japan)) and Q&A service (Yahoo! Chiebukuro) in Japan. We show the queries examples in Table 1.

- Flu patients:

The National Institute of Infectious Diseases (NIID) [29] reports data on the weekly number of patients with ILI symptoms in Japan. NIID reports information related to influenza every week through infectious disease weekly report (IDWR) [30]. These reports have a delay of about 7 days attributable to the time necessary for aggregating clinical information. We use the number of ILI patients as past flu patients as resources and the correct data for our experiments.

- Search Engine:

We use queries in Yahoo! Japan search [31], which many people use in Japan. We extract words as candidate queries: the top 50 words with the highest weights in tf-idf value and the top 100 words with the highest frequencies in tweets including "*I-N-FU-LU-E-N-ZA*" ("influenza") and "*I-N-FU-LU*" ("influ" is the abbreviation of influenza in Japan) from 2013/10/02 through 2015/05/10 is used to select queries for use. From these words, we select 13 words for which the correlation coefficients between the number of search query logs using these candidate queries and the actual number of flu patient is greater than 0.70, as queries.

- Social Networking Service:

We use posts in Twitter [32], one of social networking services. With the same method as search engine, we select queries from candidate queries. We select 18 words for which the correlation coefficient is greater than 0.75.

- On-line Shopping:

We use queries in Yahoo! Shopping [33], one of on−line shopping services. We use the different method from that of search engine because it is assumed that words different from social networking service are used when the flu is caught. We arbitrarily select 10 words related to the flu.

- Q & A Service:

We use queries in Yahoo! Chiebukuro [34], one of popular Q & A services in Japan. The same method as the queries selection in on−line shopping service is used. We arbitrarily select 9 words related to the flu.

A summary of our resources and examples of queries are represented in Table 1 and Table 2. A volume example of five resources is shown in Figure 2.

Table 1. Summary of resources: queries how to select, examples of the query and the number of features

| | Method of selection | Example of queries | Number of features |
|---|---|---|---|
| (a) Past flu patients | | | 52 |
| (b) Search query logs (Yahoo! Japan) | Select words as queries with correlation coefficient larger than 0.70 | 「高熱」(‘high fever’) 「いんふる」(‘influ’ in hiragana) | 13 |
| (c) Social media posts (Twitter) | Select words as queries with correlation coefficient larger than 0.75 | 「兄」(‘big brother’) 「辛い」(‘painful’) | 18 |
| (d) Shopping query logs (Yahoo! Shopping) | Arbitrarily select words related to flu patients | 「空気清浄機」(‘air purifier’) 「マスク」(‘mask’) | 10 |
| (e) Q&A Service logs (Yahoo! Chiebukuro) | Arbitrarily select words related to flu patients | 「インフル」(‘influ’) 「A 型インフルエンザ」(‘influenza A virus’) | 9 |

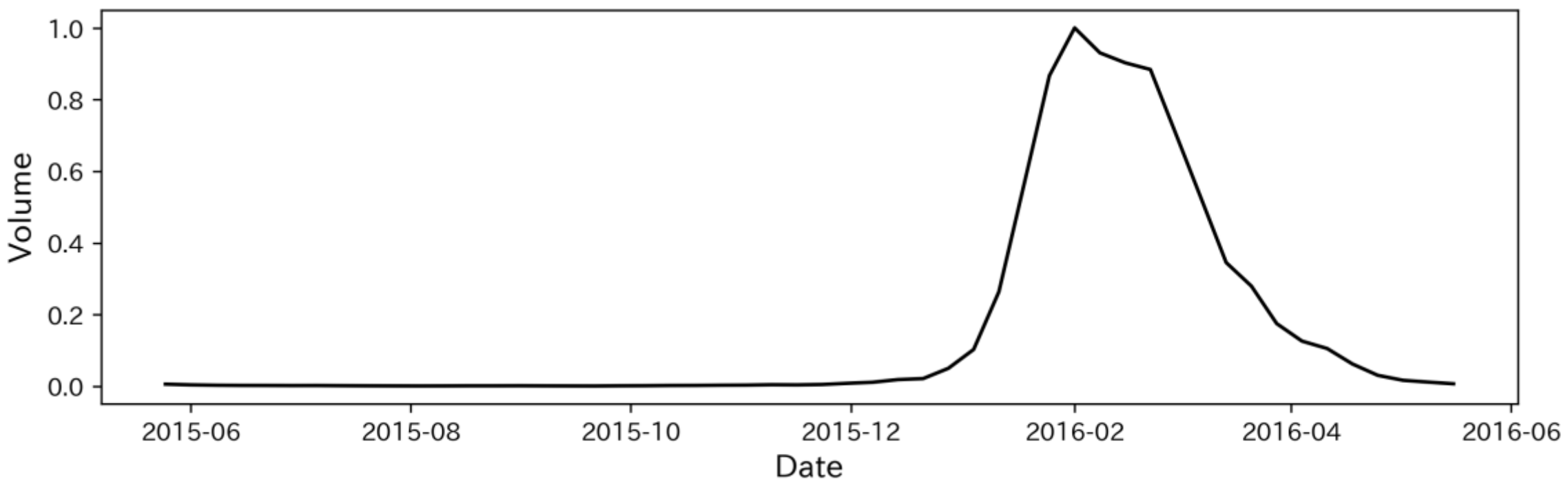

(a) Past Flu Patients

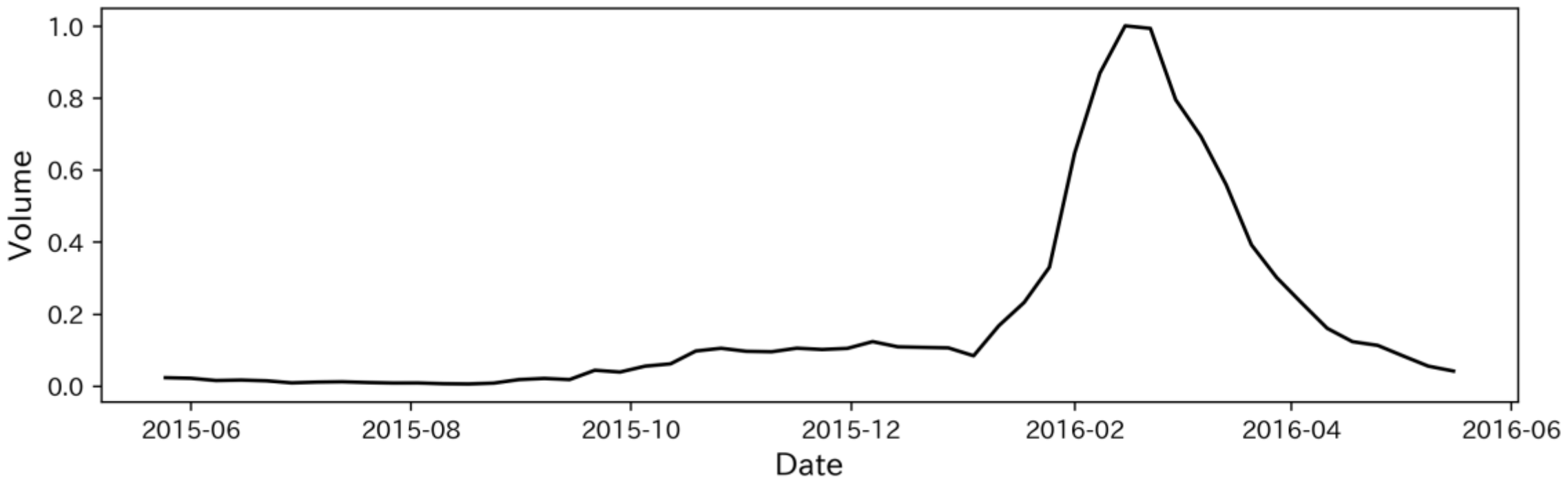

(b) Search Engine: インフル (‘influ’)

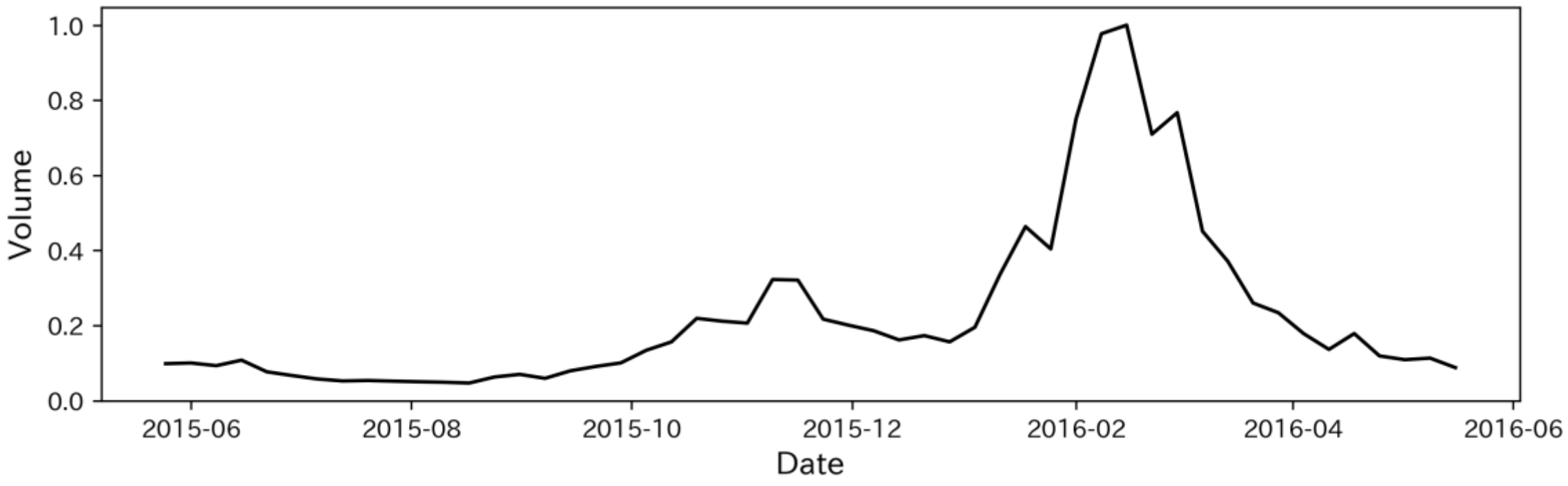

(c) SNS：インフル (‘influ’)

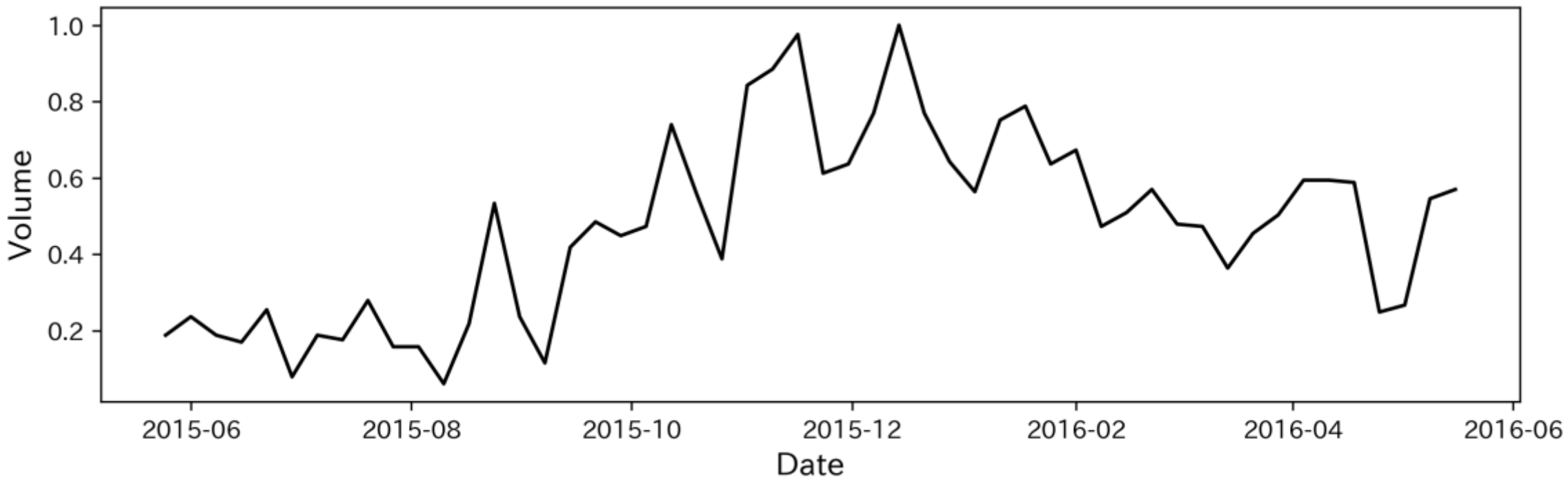

(d) On-line Shopping：風邪薬 (‘cold medicine’)

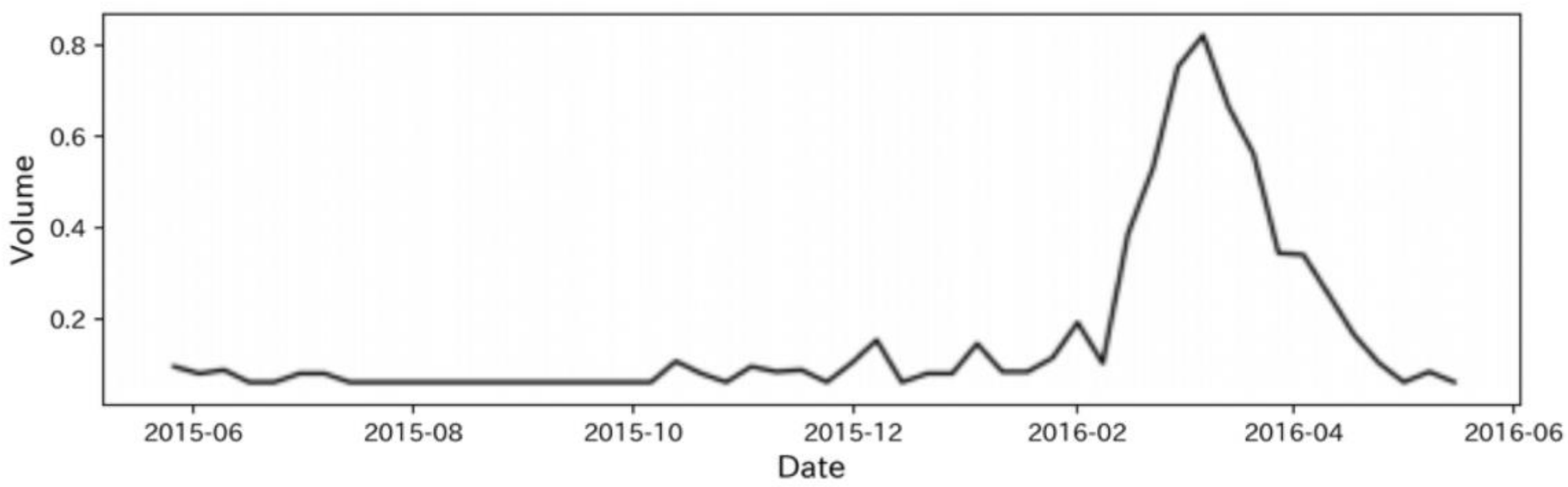

(e) Q & A Service：インフル ('influ')

Figure 2. Examples of time series about original data in each resource: (a) Past flu patients (b) Search Engine 'インフル (influ)', (c) SNS 'インフル (influ)', (d) Shopping service '風邪薬 (Cold medicine)', (e) Q&A service 'インフル (influ)'.

## Model

We selected four machine learning algorithms to apply in this study: LASSO regression, Huber regression, Support Vector Machine regression with Linear kernel (SVR), and Random Forest. Aside from Huber regression, each was chosen for simplicity. Each is known to predict flu outbreaks. Huber regression is robust against outlier response variables. We regard this as suitable for UGC data, which are extremely noisy.

- Lasso Regression:

Lasso regression, is one of regression methods, performs both variable selection and regularization by adding Lasso regularization when the parameters are fitted. This method is used in many situations because a sparse solution is obtained and is possible to select some variables from many variables.

$$\boldsymbol{y} = \boldsymbol{c}\beta \ + \ \epsilon$$
$$\beta = argmin_{\beta}(\|\boldsymbol{y} - \boldsymbol{X}\beta\|_2^2 + \lambda\|\beta\|_1)$$

$\boldsymbol{y}$ is observation vector, $\boldsymbol{X}$ is feature vector, $\beta$ is regression variable vector and $\epsilon$ error vector.

- Huber Regression:

Huber Regression is one of robust regression methods, which are strong against outliers not following other observed patterns. Huber Loss function is used in the parameter fitting, not the common squared loss function. Huber Loss function is described below:

$$a = \frac{\|y - X\beta\|_1}{\sigma}$$
$$L_\delta(a) = \begin{cases} a^2 & (|a| \le \delta) \\ 2|a| - 1 & (|a| > \delta) \end{cases}$$

1 is used as $\delta$ parameter of the function. This function has the feature to decrease loss value of outliers.

Table 2. Summary of queries, which we select in each service

| Search query logs (Yahoo! Japan) | Social media posts (Twitter) | Shopping query logs (Yahoo! Shopping) | Q&A Service logs (Yahoo! Chiebukuro) |
|---|---|---|---|
| インフル (flu in Katakana)<br>いんふる (flu in Hiragana)<br>インフルエンザ (influenza in Katakana)<br>いんふるえんざ (Influenza in Hiragana)<br>マスク (mask)<br>解熱 (fever reduction)<br>学級閉鎖 (class closure)<br>風邪をうつす (have a cold)<br>インフル 流行 (flu epidemic)<br>インフル 予防 (flu prevention)<br>インフル 薬 (flu medicine)<br>インフル 症状 (flu symptom)<br>インフル 潜在期間 (flu potential period) | インフル (flu)<br>インフルエンザ (influenza)<br>インフル 元気 (flu fine)<br>インフル 心配 (flu worry)<br>インフル 手 (flu hand)<br>インフル 体調 (flu physical condition)<br>インフル 大変だ (flu serious)<br>インフル 熱 (flu fever)<br>インフル 微熱 (flu slight fever)<br>インフル ゆっくり (flu slowly)<br>インフル 汗 (flu sweat)<br>インフル 帰る (flu go back)<br>インフル 流行 (flu epidemic)<br>インフル 辛い (flu painful)<br>インフル 兄 (flu big brother)<br>インフル 薬 (flu medicine)<br>インフル 症状 (flu symptom)<br>インフル 学校 (flu school) | インフルエンザ (influenza)<br>ハンドソープ (hand soap)<br>マスク (mask)<br>温度計 (thermometer)<br>加湿器 (humidifier)<br>空気清浄機 (air cleaner)<br>湿度計 (Hygrometer)<br>消毒液 (Antiseptic solution)<br>石鹸 (soap)<br>風邪薬 (cold medicine) | インフル (flu)<br>インフルエンザ (influenza)<br>A 型インフルエンザ (Type A influenza)<br>B 型インフルエンザ (Type B Influenza)<br>予防接種 (vaccination)<br>タミフル (tamiflu)<br>風邪 (cole)<br>熱 (fever)<br>体調 (physical condition) |

- Support Vector Machine Regression with Linear kernel (SVR):

Support Vector Machine Regression (SVR) is an extension of Support Vector Machine to regression, based on the idea of margin maximization. The method has the feature to implicitly map their inputs into feature spaces. The regression equation of SVR is described below:

$$K(x^i, x^j) = \phi(x^i)\phi(x^j)$$

$$f(x^j) = \sum_{i=1}^{n}(\alpha_i - \alpha_i^*)\,K(x^i, x^j) + c$$

$\phi(x^i)$ is the feature vector, $K(*)$ is Kernel function and $\alpha_i$, $\alpha_i^*$ are lagrange multipliers.

- Random Forest (RF):

Random Forest (RF) approach has been used in several public health studies such as the prediction of deer mouse population dynamics, not only influenza studies. This approach is a tree-based method that stratifies or segments the predictor space into several simple regions. It is frequently used to analyze variable importance.

## Experimental Settings

We use available UGC data for prediction of the number of flu patients in one week. We then use the past flu patients data from 2 weeks to 53 weeks before the predictive point because the report of flu patients in Japan is announced in about one week. Therefore, one year's flu patient data are used as features for prediction. All features are applied to standardization.

We assess the predictive performance for three flu seasons (from 2015-05-25 to 2016-05-22, from 2016-05-23 to 2017-05-21 and from 2017-05-22 to 2018-05-21), each being a year-long period. We use all data from 2013-10-02 as training data that are available prior to each data point being predicted.

## Evaluation Metrics

This report describes three evaluation metrics to compare predictive performance: the coefficient of determination $R^2$, mean absolute error (MAE), and mean absolute percent error (MAPE). Actually, $R^2$ is a measure of how well predicted values conform to true values. MAE is a measure of the average magnitude of difference between predicted and true values. MAPE is a measure of the average magnitude of the difference percentage between predicted and true values. These are defined as shown below.

$$R^2 = \frac{\sum_{t=1}^{n}(F_t - A_t)^2}{\sum_{t=1}^{n}\left(A_t - \overline{(A_t)}\right)}$$

$$\mathrm{MAE} = \frac{\sum_{t=1}^{n}|F_t - A_t|}{n}$$

$$\mathrm{MAPE} = \frac{1}{\mathrm{n}}\sum_{\mathrm{t}=1}^{\mathrm{n}}\left|\frac{\mathrm{F_t} - \mathrm{A_t}}{\mathrm{A_t}}\right| * 100\ (\%)$$

Therin, $n$ denotes the sample number, $A_t$ denotes true values, and $F_t$ represents predicted values. $R^2$ is the most important indicator in terms of flu prediction: it shows how well a model fits and the accuracy of prediction in the epidemic season.

## Results

Table 3 shows the performance of the respective methods and terms for prediction, as measured for each evaluation metric. The model under Huber Regression had the best performance each year in terms of $R^2$ and $\mathrm{MAE}$. However, in terms of $\mathrm{MAPE}$, the model under Random Forest showed the best performance each year. Figure 3 presents plots of predicted flu values from 2016–05 through 2017–05 using four methods. Given the nature of evaluation metrics, the model under Huber Regression outperforms other models, especially during the epidemic period (December–March). The model under Random Forest achieves stable prediction irrespective of the epidemic period or non–epidemic period. Each model and each feature based on this result are assessed below.

Huber Regression, which achieves the highest accuracy for two metrics, is a modified model in terms of the loss function. It is therefore robust against outlier objective variables. This model also has some degree of robustness against outlier explanatory variables. These experiments use outlying observation values in some search query logs and zero values for a period of time because of bad Twitter crawling conditions that occur as a result of the use of many resources. These difficulties actually lead to good results of the model under Huber Regression. Some difficulties such as sharply increased numbers of accesses and change in crawling are always present. Therefore, using Huber Regression to predict infectious diseases such as influenza is useful in practice.

Prediction using Random Forest has higher accuracy than Huber Regression as measured by $\mathrm{MAPE}$. In temperate areas such as Japan, flu patients during non–epidemic terms are around 1,000. When one variable changes slightly when using the method of regression, the prediction value can vary greatly. $\mathrm{MAPE}$ similarity reflects large value changes. Consequently, the model under Random Forest outputs stable values irrespective of non–epidemic terms and gives good results in terms of $\mathrm{MAPE}$.

The model under Huber Regression is more important and suitable for practical use than under Random Forest because the epidemic prediction term is more important than the non–epidemic term.

Table 3. Accuracy of four models used for flu prediction

| | 2015/05 – 2016/05 | | | 2016/05 – 2017/05 | | | 2017/05 – 2018/05 | | |
|---|---|---|---|---|---|---|---|---|---|
| | $R^2$ | MAE | MAPE | $R^2$ | MAE | MAPE | $R^2$ | MAE | MAPE |
| Lasso Regression | 0.488 | 30481.20 | 3004.88 | 0.676 | 19138.73 | 1801.58 | 0.572 | 24251.58 | 433.45 |
| Huber Regression | **0.907** | **8457.69** | 242.54 | **0.889** | **6359.84** | 85.93 | **0.917** | **7949.70** | 42.94 |
| SVR (Linear kernel) | 0.693 | 15083.25 | 199.74 | 0.887 | 12113.98 | **76.17** | 0.619 | 22192.46 | 116.68 |
| Random Forest | 0.718 | 13857.41 | **105.23** | 0.864 | 10628.23 | 117.93 | 0.823 | 12240.59 | **39.19** |

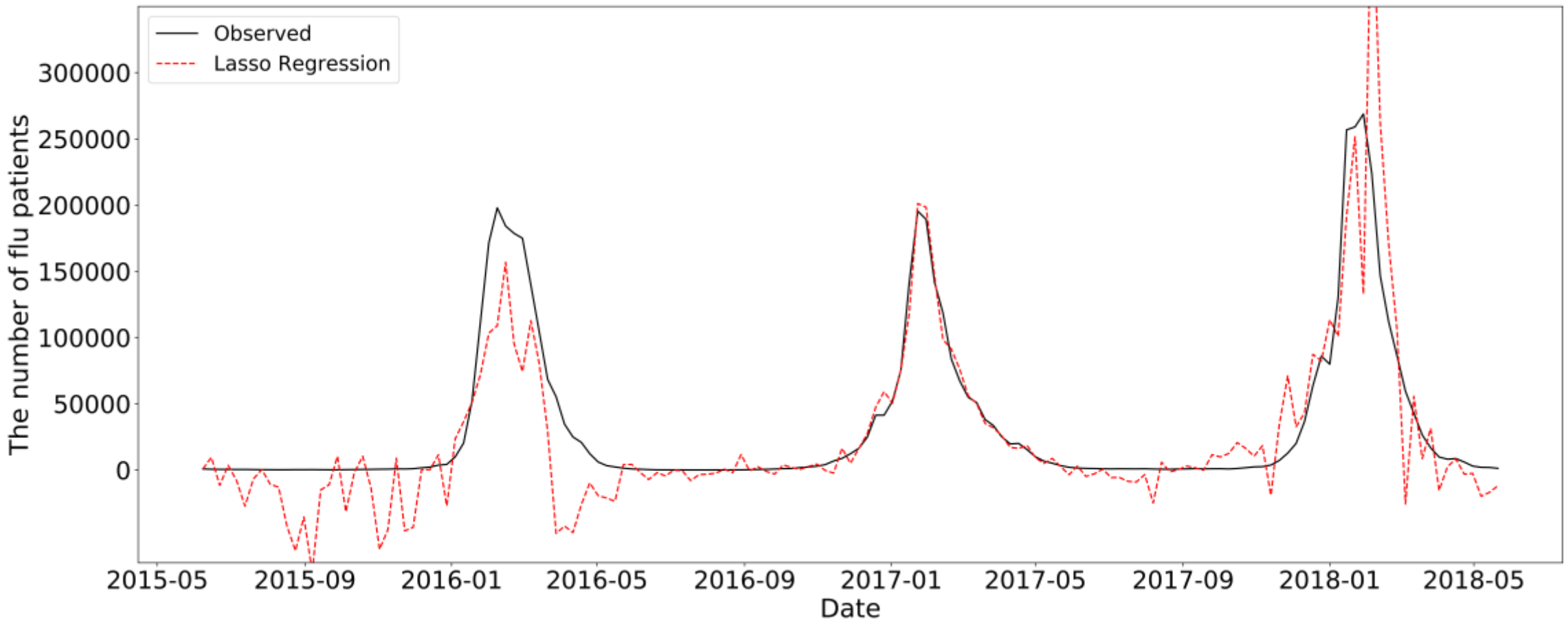

(a) Lasso regression

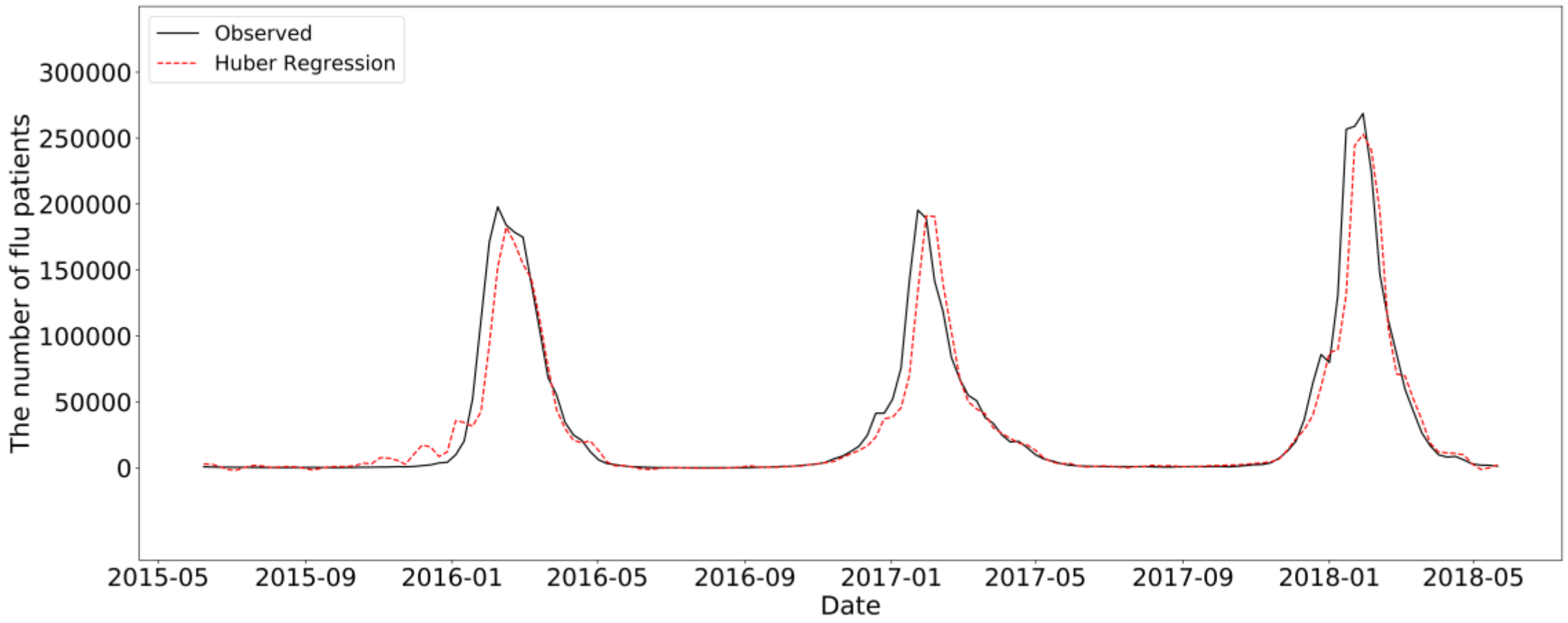

(b) Huber regression

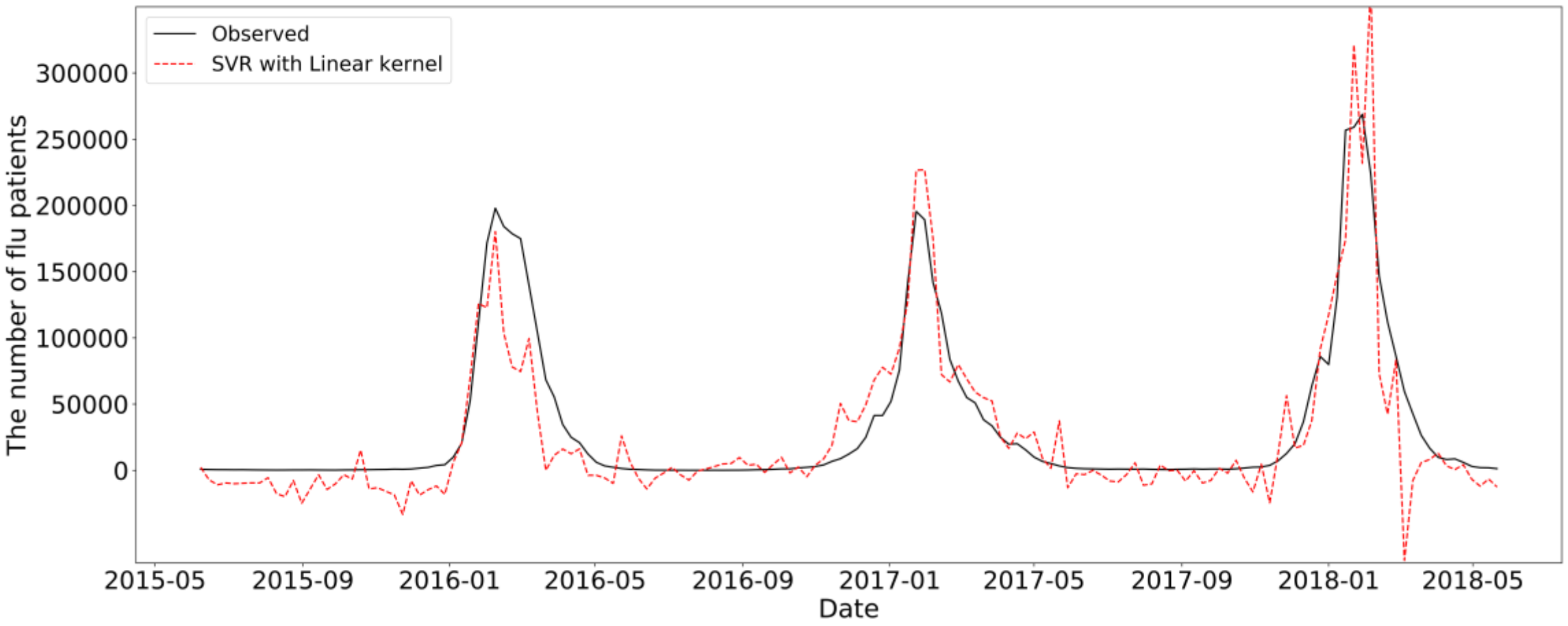

(c) SVR

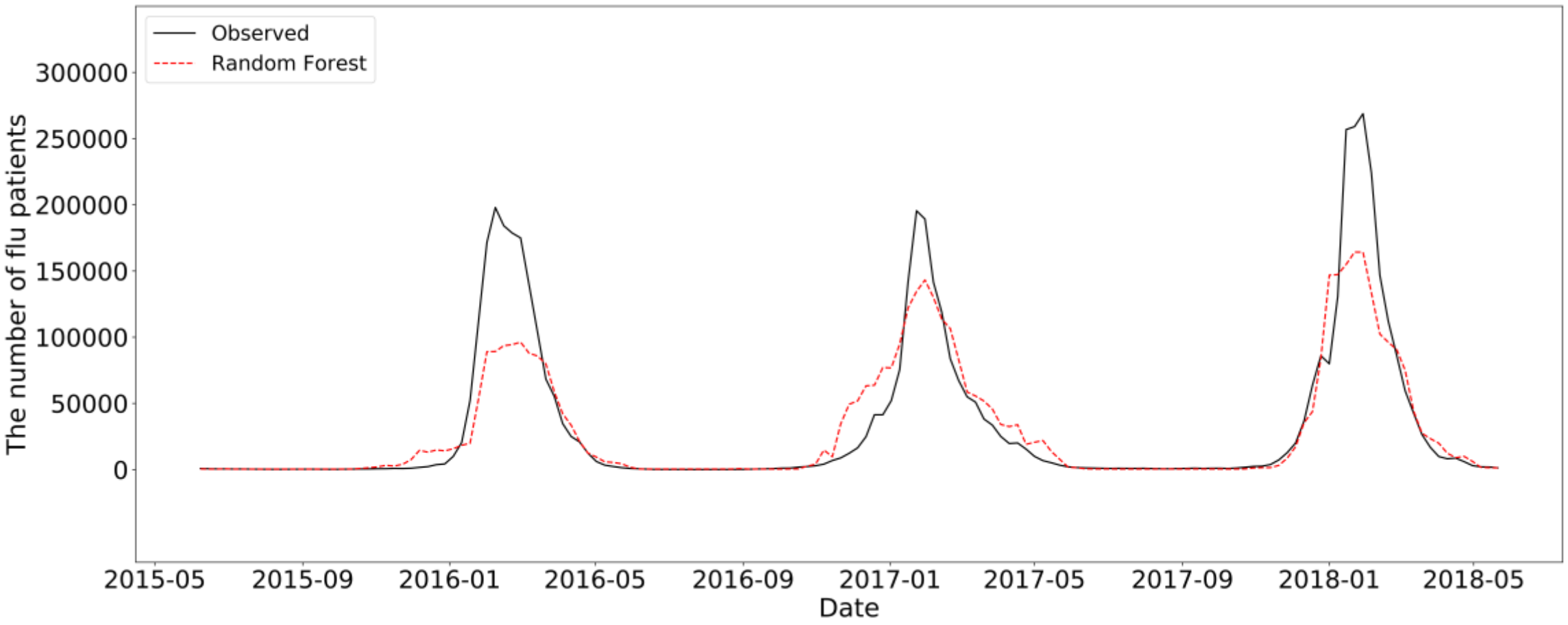

(d) Random Forest

Figure 3. Comparative flu prediction plot of four models: (a) Lasso regression, (b) Huber Regression, (c) SVR, (d) Random Forest

## Discussion

### Effect of respective features

Table 4 shows the number of flu patients predicted using the model under Huber Regression with all resources except one resource to validate respective resources contributions. The model with all resources except Twitter resource achieves higher accuracy than all resources for 2017–05 through 2018–05 because tweet crawling does not work well during this period.

Table 4 also shows different degrees of effectiveness among years and features, except for past flu patients. Results show that making a consistently useful model for predicting flu outbreaks is difficult. However, results show that the past numbers of flu patients are important predictive data.

Table 4. Study of accurate on resource comparison under Huber Regression
* failure of crawling in a certain period during 2017 and 2018

| | 2015/05 – 2016/05 | | | 2016/05 – 2017/05 | | | 2017/05 – 2018/05 | | |
|---|---|---|---|---|---|---|---|---|---|
| | $R^2$ | MAE | MAPE | $R^2$ | MAE | MAPE | $R^2$ | MAE | MAPE |
| All | **0.907** | **8457.69** | **242.54** | 0.889 | **6359.84** | 85.93 | 0.917 | 7949.70 | 42.94 |
| w/o Search query logs | 0.906 | 8637.29 | 254.78 | 0.903 | 6419.06 | 110.41 | 0.900 | 8439.72 | 45.59 |
| w/o Social media posts | **0.907** | 8573.25 | 263.66 | 0.880 | 6754.79 | 59.62 | **0.921*** | **7303.47*** | **32.32*** |
| w/o Shopping query logs | 0.906 | 8603.04 | 255.19 | **0.904** | 6887.97 | **110.09** | 0.902 | 8258.65 | 37.90 |
| w/o Q&A service logs | 0.906 | 8569.50 | 249.81 | 0.881 | 6443.73 | **73.03** | 0.910 | 8024.62 | 37.75 |
| w/o Flu patients in the past | 0.489 | 19065.75 | 302.54 | 0.488 | 18917.90 | 710.61 | 0.449 | 25941.20 | 173.73 |

Table 5. Accuracy of the flu prediction under ARIMA model

| | 2015/05 – 2016/05 | | | 2016/05 – 2017/05 | | | 2017/05 – 2018/05 | | |
|---|---|---|---|---|---|---|---|---|---|
| | $R^2$ | MAE | MAPE | $R^2$ | MAE | MAPE | $R^2$ | MAE | MAPE |
| ARIMA (3,1,2) | 0.823 | 12954.98 | 57.77 | 0.723 | 11472.42 | 131.01 | 0.733 | 16250.42 | 61.69 |

Table 5 presents prediction results for the number of flu patients using the Autoregressive integrated moving average (ARIMA) model with only the number of flu patients as a resource. Results from Table 5 demonstrate that sufficient accuracy for influenza prediction can be achieved using only the past number of flu patients. The model with all resources achieved better results than those using only one resource, but it is not certain whether the cost of using all resources to produce a marginally better model in some cases is appropriate for improved prediction accuracy overall.

From these results, we infer that social media posts and search query logs are useful for prediction in any country with unstable sources of public health information (e.g. past flu patients), such as economically developing countries. However, these resources present fewer opportunities (such as flu prediction time lag modification) for use in countries with public health information related to annual cycle infection.

## Change point analysis

To ascertain whether the trend of time series on any resources is reflected in the trend of flu patient count, or not, we use Bayesian change point method [26,27]. Bayesian change point analysis can be used to detect signals related to subtle changes within time series data. This method is used widely for various fields: interest rate data, cancer-related gene expression data [27], and some resource data (Wikipedia, Twitter, search query) for flu detection [5]. This method calculates change point probabilities $p \in [0,1]$ of each point $i$ on the basis of the hypothesis that each observation point is independent. With the partition of time series $\rho = (U_1, U_2 \ldots, U_n)$, where n is the number of observations and $U_i = 1$ indicates a change point at position at $i+1$. The transition probability $p$ at the position $i+1$ is obtained from below:

$$\frac{p_i}{1-p_i} = \frac{P\left(U_i = 1 | \boldsymbol{X}, U_j, j \neq i\right)}{P\left(U_i = 0 | \boldsymbol{X}, U_j, j \neq i\right)}$$

$$= \frac{\left[\int_0^\gamma p^c (1-p)^{n-b-1} dp\right]\left[\int_0^\lambda \frac{w^{b/2}}{(W_1 + B_1 w)^{(n-1)/2}} dw\right]}{\left[\int_0^\gamma p^{b-1} (1-p)^{n-b} dp\right]\left[\int_0^\lambda \frac{w^{(b-1)/2}}{(W_0 + B_0 w)^{(n-1)/2}} dw\right]}$$

$W_0$ and $W_1$ are the within blocs sums of squares obtained when $U_i = 1$ and $U_i = 0$, respectively. $B_0$ and $B_1$ are the between blocs sums of squares obtained. $\boldsymbol{X}$ is the data. $\lambda$ and $\gamma$ are hyper-parameters, which are taken from [0,1]. $j$ indicates ending position and $b$ is the number of blocks in the partition. MCMC updates the various sums of squares ($W_0$, $W_1$, $B_0$ and $B_1$) in each step.

We use the R package "bcp" ver. 4.0.2 [26], which implements Bayesian change point analysis using a complex Markov Chain Monte Carlo (MCMC) approximation. With a view to ascertaining whether change points in flu patients match those of internet-based resources, we regard any point for which the probability of a change occurring exceeds 50% as a change point. Change points of web-based resources that occurred from 1 week before to 1 week after a change point of the time series for flu patients are regarded as true matching change points. We calculate Sensitivity and the positive predictive value (PPV) to validate the resource effectiveness. This calculation uses matching change points (true positive) and change points detected for flu patient data, but not for web-based resources (true negative). It also uses change points detected for web-based resources but not for flu patient data (false positive). Sensitivity and PPV are defined as shown below.

$$Sensitivity = \frac{true\ positive}{true\ positive + false\ negative} (\%)$$

$$PPV = \frac{true\ positive}{true\ positive + false\ positive} (\%)$$

For the change point analysis, we set the number of MCMC iterations as 500 and hyper-parameter $\lambda$ and $\gamma$ as 0.1. We then select three queries, which are the highest correlation coefficient between each query and the ILI rate, from each resource.

Results are presented in Table 6. Two metrics about social media posts are bad because of the unstable amount of tweets between 2016-05 and 2017-05. However, this high value of two metrics related to search query logs and social media posts for three years represents these resources as a good predictor. Presumably, two resources are stable contributions to flu prediction: Search query logs and Social media posts.

For shopping query logs, *PPV* is high but *Sensitivity* is low because this time series reflect few behaviors; they capture little change points of flu patients. For Q&A service logs, *Sensitivity* is high but *PPV* is low. This result demonstrates that the time series related to Q&A service logs capture the change points for flu patients. However, results capture the flatness of the times series as change points because Q&A service logs include much noise.

Table 6. Sensitivity and PPV for four resources

| | 2015/05 – 2016/16 | | 2016/05 – 2017/05 | | 2017/05 – 2018/05 | |
|---|---|---|---|---|---|---|
| | Sensitivity (%) | PPV (%) | Sensitivity (%) | PPV (%) | Sensitivity (%) | PPV (%) |
| Search query logs | 60 | 43 | 73 | 57 | 80 | 75 |
| Social media posts | 60 | 70 | 8 | 33 | 75 | 71 |
| Shopping query logs | 38 | 56 | 25 | 60 | 47 | 78 |
| Q&A service logs | 55 | 32 | 31 | 25 | 47 | 57 |

## Conclusions

Our research proposed a model for flu prediction in Japan using five resources simultaneously and verified the usefulness of each resource. The present results suggest the following:

- Huber Regression is strong for outliers and is suitable for the flu prediction using UGC contents.
- The ensemble model combining information from multiple independent resources is superior to choosing some resources as predictors. However, its degree of increased accuracy is slight.
- The number of past flu patients is a good predictor.

- Regarding resource characteristics, shopping query logs show few behaviors; Q&A service logs have much noise.

Some earlier studies include discussions of useful resources such as search queries or Twitter [5, 16]. It appears difficult to conclude which one is better. Results presented herein suggest that the usefulness of respective resources for flu prediction depends on the year and shows no great difference. Our ensemble model produces reliable, robust, and accurate estimates. However, some models that do not include one of the resources show better scores for some years because of crawling malfunctions. In some countries with sufficient public health information, the use of a single UGC (search query or Twitter) in addition to past flu patient data can produce a highly accurate prediction model. On the other hand, for countries with insufficient public health organizations and information, the UGC is expected to have a proportionally higher value as a flu prediction model component.

## Conflicts of Interest

The authors have no conflicts of interest directly relevant to the content of this article.